\documentclass[useAMS,usenatbib]{mn2e}
\usepackage{graphicx}
\usepackage{amssymb}
\usepackage{amsmath}
\usepackage{algpseudocode}
\usepackage{multicol}
\usepackage{longtable}
\usepackage{algorithm}
\usepackage{cprotect}

\title[Merging]{Dark-matter halo mergers as a fertile environment for low-mass Population III star formation} 

\author[S. Bovino et al.]{ S. Bovino\thanks{Corresponding author: sbovino@astro.physik.uni-goettingen.de}$^1$, M. A. Latif$^1$, T. Grassi$^{2,3}$, and D. R. G. Schleicher$^1$\\
$^1$Institut f\"ur Astrophysik Georg-August-Universit\"at, Friedrich-Hund Platz 1, D-37077 G\"ottingen, Germany\\
$^2$Centre for Star and Planet Formation, Natural History Museum of Denmark, \O ster Voldgade 5-7, DK-1350 Copenhagen, Denmark\\
$^3$Niels Bohr Institute, University of Copenhagen, Juliane Maries Vej 30, DK-2100 Copenhagen, Denmark}
\begin{document}
\newcommand{\ith}{$i$th }
\newcommand{\jth}{$j$th }
\newcommand{\nth}{$n$th }

\newcommand{\dd}{\mathrm d}
\newcommand{\mA}{\mathrm A}
\newcommand{\mB}{\mathrm B}
\newcommand{\mC}{\mathrm C}
\newcommand{\mD}{\mathrm D}
\newcommand{\mE}{\mathrm E}
\newcommand{\mH}{\mathrm H}
\newcommand{\mSi}{\mathrm Si}
\newcommand{\mO}{\mathrm O}
\newcommand{\cmc}{\mathrm{cm}^{-3}}
\newcommand{\real}{\mathbb R}
\newcommand{\superscript}[1]{\ensuremath{^{\scriptscriptstyle\textrm{#1}\,}}}
\newcommand{\trader}{\superscript{\textregistered}}

\newcommand\mnras{MNRAS}
\newcommand\apj{ApJ}
\newcommand\aap{A\&A}
\newcommand\apss{Ap\&SS}

\date{Accepted *****. Received *****; in original form ******}

\pagerange{\pageref{firstpage}--\pageref{lastpage}} \pubyear{2013}

\maketitle

\label{firstpage}

\begin{abstract}
While Population III stars are typically thought to be massive, pathways towards lower-mass Pop III stars may exist when the cooling of the gas is particularly enhanced.
A possible route is enhanced HD cooling during the merging of dark-matter halos.
The mergers can lead to a high ionization degree catalysing the formation of HD molecules and may cool the gas down to the cosmic microwave background (CMB) temperature.
In this paper, we investigate the merging of mini-halos with masses of a few 10$^5$~M$_\odot$
and explore the feasibility of this scenario.
We have performed three-dimensional cosmological hydrodynamics calculations with the \verb|ENZO| code, solving the thermal and chemical evolution of the gas by employing the astrochemistry  package \verb|KROME|. 
Our results show that the HD abundance is increased by two orders of magnitude compared to the no-merging case and the halo cools down to $\sim$60 K triggering fragmentation.
Based on Jeans estimates the expected stellar masses are about 10 M$_\odot$.
Our findings show that the merging scenario is a potential pathway for the formation of low-mass stars.

\end{abstract}


\section{Introduction}\label{introduction}
The formation modes of the first generations of stars have been a challenging subject for many years and today still represent a central topic of research in modern cosmology. 
The current understanding is that these stars were formed in mini-halos of about 10$^6$ M$_\odot$ around $z$~=~20-30, which is strongly supported by many numerical simulations performed over the years \citep{Abel2002,Bromm2002,Ciardi05,Yoshida2006,Oshea2007,Greif2008,Greif2012}. Since \citet{Saslaw1967}, it is widely accepted that the collapse process leading to the formation of these early objects is mainly driven by H$_2$ cooling \citep{Peebles1968,Palla1983}. Due to the lack of a permanent dipole moment the hydrogen molecule becomes an ineffective coolant below $\sim$200 K when it reaches the local thermodynamic equilibrium (LTE). Considering these thermal conditions, it has been postulated that the final mass of the first generation of stars should be of the order of $\sim$100~M$_\odot$ or more leading to a resultant short life \citep{Abel2002,Bromm2002,BrommRev2013}. However, recent numerical simulations \citep{ClarkGlover2011,Greif2012,Stacy2012} suggest that the characteristic mass scale of the first stars should be lowered to several ten solar masses, if fragmentation occurs.
The large parameter study by \citet{Hirano2013} indicates a significant spread around this mass scale, and extreme values up to 1000 M$_\odot$.

Comprehending the conditions under which the first stars were formed is very important to study the early history of the Universe as they initiated cosmic reionization and the enrichment of the intergalactic medium \citep[e.g.][]{Barkana2001,Schneider06,Schleicher08}, and may also have seeded the formation of supermassive black holes \citep{Li2007,Volontieri2012,LatifBH,Hosokawa2013,LatifBH2}.

Many authors \citep{Bromm2002,Johnson2006MNRAS.366..247J,Yoshida2006,Ripamonti2007,Greif2008,McGreer2008} also discussed the possible role of HD as an additional coolant which may have led to the formation of lower-mass stars. They found that in the presence of a higher electron abundance, HD cooling can become efficient allowing the gas to reach the temperature of the CMB. The need of a high ionization fraction suggested the introduction of a second mode to form primordial stars named Pop~III.2 to be distinguished from Pop~III.1 in which H$_2$ cooling dominates throughout \citep[see][]{Tan2008,BrommRev2013}. The Pop III.2 stars are expected to be assembled from a gas which is still metal-free but embedded in ionized environments. The latter may be triggered by the explosions of the first supernovae or HII regions or strong shocks \citep[e.g.][]{Mackey2003,Yoshida2007,Greif2008}. 
Lower-mass stars may also be formed within the Pop III.1 paradigm as shown by \citet{Hirano2013}. They suggested that in a slowly collapsing cloud the compressional heating is reduced and may create the conditions for the HD cooling to become effective. In some of the above mentioned papers \citep{Bromm2002,Ripamonti2007,McGreer2008} the authors claimed that in mini-halos of a few $\sim$10$^5$~M$_\odot$ HD can efficiently be formed and overcome the H$_2$ cooling rate if the temperature drops below $\sim$200~K at densities of at least 10$^4$ cm$^{-3}$.
\citet{Stacy2013} performed cosmological simulations of the formation and growth of Pop III stellar systems and argued that a mini-halo formed at redshift 15 and supported by high baryonic angular momentum can lead to the formation of 1-5 M$_\odot$ stars even in the absence of HD cooling. Lower-mass stars are thus also included in the expected scatter of the Pop III.1 distribution.

\citet{McGreer2008} concluded that in halos with masses $>$10$^6$ M$_\odot$ the cooling is dominated by H$_2$ and the final stars could have a mass of $\sim$100~M$_\odot$, while for lower mass halos HD cooling becomes important and leads to stellar masses between 10-40~M$_\odot$. However, the occurrence of such low-mass systems was found to be less frequent. These conclusions are in overall agreement with the 1D Lagrangian simulation reported by \citet{Ripamonti2007} and with the robust statistical study presented by \citet{Hirano2013}.

The main idea behind the possible role of HD in primordial star formation comes from the series of chemical reactions which leads to the formation of HD. This is strongly related to the ionization fraction because the main formation path for HD is
\begin{equation}
	\mathrm{H_2 + D^+ \rightarrow HD + H^+}\\,
\end{equation}
as largely discussed in earlier papers \citep{Stancil1998,Galli2002,Stancil2011}. The only way to trigger the formation of HD is then to boost the formation of H$_2$ pumping the main reactive channel which is directly linked to H$^-$ via the following reactions:
\begin{eqnarray} 
        &\mathrm{H} + \mathrm{e^-} \rightarrow \mathrm{H^-} + \gamma\\
	&\mathrm{H^-} + \mathrm{H} \rightarrow \mathrm{H_2} + \mathrm{e^-}. 
\end{eqnarray} 
In addition the formation and destruction of D$^+$ should be considered
\begin{equation} 
        \mathrm{D^+} + \mathrm{H} \rightleftharpoons \mathrm{H^+} + \mathrm{D}.
\end{equation} 
This leads us to the conclusion that a high ionization fraction boosts the formation of HD via a chain of chemical events that includes the key species in the following way $\mathrm{e^-\rightarrow H^-\rightarrow H_2 \rightarrow HD}$. In addition, we note that also an increase of the gas temperature will boost the formation of H$^{-}$, and thus H$_2$ and HD.

Another possible route for primordial star formation (Pop III.2) which also involves the HD cooling has been proposed by \citet{Merging2006}, and \citet{Prieto2012,Prieto2013}. The idea suggested by \citet{Merging2006} is based on the merging of dark-matter halos within the context of the hierarchical scenario of structure formation \citep{Barkana2001,Ciardi05}. Merging induces compression and the consequent formation of shock-waves which may enrich the baryonic component of the gas by free electrons catalyzing the HD formation. The authors consider a simplified one-dimensional model where two identical halos collide and follow the non-equilibrium evolution of the HD behind the shock-waves which formed during the merging. In these halos, they further assume a uniform density equal to the virial density.
They found that to have a higher ionization degree and to boost the formation of HD the merger masses should be $M>8\times 10^6[(1+z)/20]^{-2}$~M$_\odot$.
In a real merger, we however expect a more complex density structure, and the strength of the shocks is likely enhanced in the central regions. Their mass scale can therefore just serve as a rough estimate.
In a recent study, \citet{Prieto2012,Prieto2013}  performed cosmological N-body simulations and computed the statistic of the halos fulfilling the \citet{Merging2006} criterion. Their findings suggest that the fraction of halos going through this phase is significant.  

In this paper we investigate the role of HD as a coolant in post-shocked environments generated by dark-matter halo merging. In particular we consider a merger of more than two halos with masses well below the threshold set by \citet{Merging2006} and run self-consistent cosmological simulations, following both the merging and the collapsing process with enough accuracy to capture important chemo-dynamical features. 

The paper is organized as follow: we first introduce the methodology and the chemical model involved in the cosmological simulations, then discuss the main results, and finally give our conclusions.

\section{Methodology}\label{sec:methodology}
To follow the merger and collapse of a halo from cosmological initial conditions
we use the cosmological hydrodynamics code \verb|ENZO|, version 2.3 \citep{Enzo2013}. \verb|ENZO| is based on an adaptive
mesh refinement (AMR) method and it is well-tested in the framework of cosmological simulations of primordial star formation \citep{Abel2002,Oshea2007,TurkScience2009,Latif2013}. It includes the split 3rd-order piece-wise parabolic (PPM) method for solving the hydrodynamical equations, while the dark matter component is modeled using the particle-mesh technique. Self-gravity is calculated via a multigrid Poisson solver.

Our approach is divided in different steps. We first run a dark-matter only low-resolution simulation to study the evolution and the merging history of the halos. A single box of 1.0 comoving Mpc with a top grid resolution of 128$^3$ cells is employed. The parameters for creating the initial conditions and the distribution of baryonic and dark matter components are taken from the WMAP seven year data \citep{Jarosik2011}. We select the most massive mini-halo with $\sim$7$\times10^5$M$_{\odot}$ at $z=12$ which is formed through merging of few times 10$^{5}$~M$_{\odot}$ halos. The progenitors were found by employing the merger tree algorithm implemented in the $\mathrm{YT}$ package \citep{Turk2011a} which makes use of the HOP halo finder \citep{Eisenstein1999}. The merger history of this halo is depicted in Fig. \ref{fig:figure1} and is very different from the isolated case which has not gone through a major merger and is formed at $z=20$. 
We re-run our simulation with the box centered in the selected mini-halo adding two additional nested grids.
As we want to follow the merging process leading to our mini-halo we adopt a refinement strategy which copes with 
the computational demands of these calculations. From $z=99$ to $z=12$ (the time at which the merging processes is almost over)
we allow 21 levels of refinement and 32 cells per Jeans length. With the same resolution we then follow
the evolution of the mini-halo until a density of 10$^8$~cm$^{-3}$, the range of interest to study the effect of HD cooling.
Our refinement strategy is based on overdensity, Jeans length, and particle mass and is applied during the course of the simulations
to ensure that all physical processes like shock waves and Truelove criterion \citep{Truelove1997} are well resolved. A final resolution of $\sim$190 AU in comoving is achieved. 
\begin{table}
	\caption{Chemical network employed in the simulations.}
	\begin{center}
		\begin{tabular}{ll}
			\hline
			\hline
			Reactions\\ 
			\hline
			\hline
                        1) H + e$^-$ $\rightarrow$ H$^+$ + 2e$^-$ & 15) H$^-$ + H $\rightarrow$ 2H + e$^-$\\
			2) H$^+$ + e$^-$ $\rightarrow$ H + $\gamma$ & 16) H$^-$ + H$^+$ $\rightarrow$ 2H\\
			3) He + e$^-$ $\rightarrow$ He$^+$ + 2e$^-$ & 17) H$^-$ + H$^+$ $\rightarrow$ H$_2^+$ + e$^-$\\
			4) He$^+$ + e$^-$ $\rightarrow$ He + $\gamma$ & 18) H$_2^+$ + e$^-$ $\rightarrow$ 2H\\
			5) He$^+$ + e$^-$ $\rightarrow$ He$^{2+}$ + 2e$^-$ & 19) H$_2^+$ + H$^-$ $\rightarrow$ H + H$_2$\\
			6) He$^{2+}$ + e$^-$ $\rightarrow$ He$^+$ + $\gamma$ & 20) 3H $\rightarrow$ H$_2$ + H\\
			7) H + e$^-$ $\rightarrow$ H$^-$ + $\gamma$ & 21) H$_2$ + 2H $\rightarrow$ 2H$_2$\\
			8) H$^-$ + H $\rightarrow$ H$_2$ + e$^-$ & 22) H$^+$ + D $\rightarrow$ H + D$^+$\\
			9) H + H$^+$ $\rightarrow$ H$_2^+$ + $\gamma$ & 23) H + D$^+$ $\rightarrow$ H$^+$ + D\\
			10) H$_2^+$ + H $\rightarrow$ H$_2$ + H$^+$ & 24) H$_2$ + D$^+$ $\rightarrow$ HD + H$^+$\\
			11) H$_2$ + H$^+$ $\rightarrow$ H$_2^+$ + H & 25) HD + H$^+$ $\rightarrow$ H$_2$ + D$^+$\\
			12) H$_2$ + e$^-$ $\rightarrow$ 2H + e$^-$ & 26) H$_2$ + D $\rightarrow$ HD + H\\
			13) H$_2$ + H $\rightarrow$ 3H & 27) HD + H $\rightarrow$ H$_2$ + D\\
			14) H$^-$ + e$^-$ $\rightarrow$ H + 2e$^-$ & 28) D + H$^-$ $\rightarrow$ HD + e$^-$\\
			\hline
		\end{tabular}
	\end{center}
	\label{tab:reactions}
\end{table}

\subsection{Chemical model}\label{sec:chemistry}
The chemical and thermal evolution of the gas is consistently solved along with the hydrodynamical equations by employing the astrochemistry package \verb|KROME| \citep{Grassi2013krome}.
The latter is well suited for the study of many chemical environments and employs very accurate and efficient solvers as already shown and discussed in previous works \citep{Bovino2013MNRAS,Grassi2013,Bovino2014}. To include \verb|KROME| in \verb|ENZO| we used the patch directly provided by \verb|KROME| and publicly released with the package\footnote{Webpage KROME: http://kromepackage.org}.
Our chemical network includes 28 reactions, listed in table \ref{tab:reactions}, and 12 species: H, H$^+$, H$^-$, H$_2$, H$_2^+$, He, He$^+$, He$^{2+}$, e$^-$, D, D$^+$, and HD. The reaction rates are the same already discussed in many papers \citep[for details see][]{Abel97,Ripamonti2007}. The network (react\_primordial) is available with \verb|KROME|. In a previous work, \citet{Turk2011ApJ} discussed the uncertainty in the H$_2$ three-body formation rates: we employ here the latest accurate available rate from \citet{Bala2013ApJ,Bala2013PhRvA} which has already been tested in primordial star formation studies in a recent paper \citep{Bovino2014}.
Heating and cooling include: H$_2$ formation heating as described in \citet{Omukai2005}, H$_2$ cooling as reported in \citet{GloverAbel08}, bremmsstrahlung, H and He line cooling \citep{Cen92}, HD cooling by \citet{Lipovka2005}, and collisionally induced emission (CIE) cooling as discussed in \citet{Grassi2013krome}. In addition, we include the optically thick correction for the H$_2$ cooling following \citet{Ripamonti2004} and \citet{Grassi2013krome}. More details about the cooling and the heating functions can be found in the \verb|KROME| paper \citep{Grassi2013krome} where they are extensively discussed.
It is worth noting that three-body processes and the CIE cooling, as well as the optical thick correction, are not important in this study as we look at an intermediate range of densities, 10$^4<n<$10$^8$ cm$^{-3}$, while the above contributions play an important role in the high-density regime ($n>10^8$ cm$^{-3}$). We decided to include them anyway for the sake of completeness.

To accurately solve the evolution of the chemical species and the temperature, we employ the following tolerances for the \verb|DLSODES| solver: a relative tolerance RTOL=10$^{-4}$, and an absolute tolerance ATOL=10$^{-10}$. These tolerances ensure good performance without loosing accuracy.

\section{Results}\label{sec:results}
\begin{figure}
\begin{center}
	\includegraphics[width=.5\textwidth]{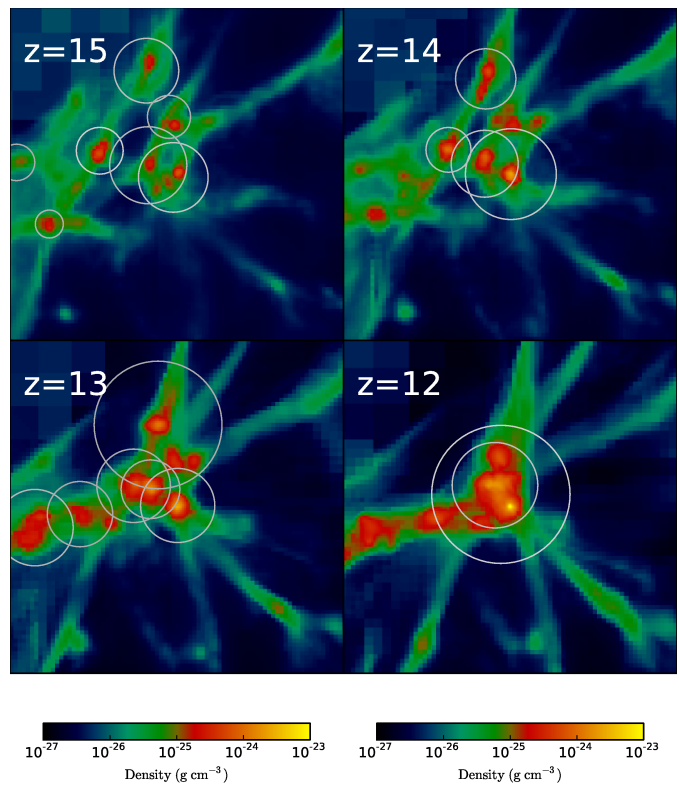}
	\caption{Density projections showing the merging of various halos at different redshifts. The overplotted circles represent halos of at least few 10$^5$ M$_\odot$ and the plot scale is 2 kpc.}\label{fig:figure1}
\end{center}
\end{figure}

\begin{figure}
\begin{center}
	\includegraphics[width=.5\textwidth]{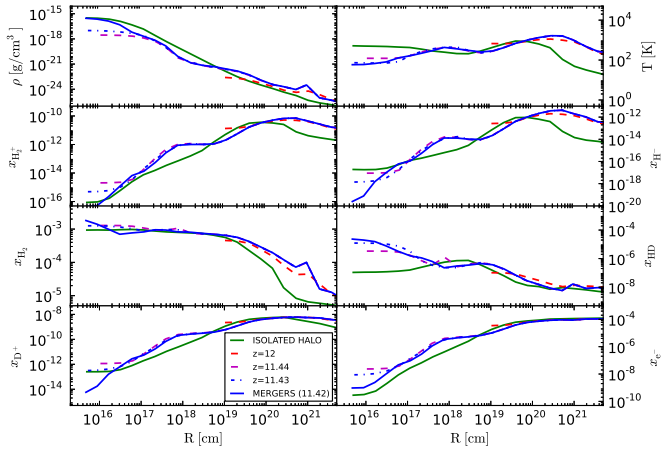}
	\caption{Radially averaged profiles for density, temperature, H$^-$, H$_2^+$, H$_2$, HD, D$^+$, and electron mass fractions, for different redshifts. The comparison between the isolated run (green solid) and the mergers (blue solid) is made taking the data at the same density peak in the simulations, i.e. at a redshift of 11.42 for the mergers case. See text for details.}\label{fig:figure2}
\end{center}
\end{figure}

In Fig. \ref{fig:figure1}, we show the merging of halos with more than 10$^5$~M$_\odot$ at different redshifts ranging from $z=15-12$. The merging process is initiated around redshift $z=15$ where several halos of a few 10$^5$ solar masses start to interact. The process gradually continues along the gas filaments and 4-5 halos converge towards the central region until $z=12$, where we clearly see two mini-halos, as depicted by the circles. The latter merge to form a mini-halo of 7$\times$10$^5$ M$_\odot$. Note here that the size of the circles corresponds to the physical mass of the halos and is provided by the halo finder employed in $\mathrm{YT}$; bigger circles represent more massive halos. We follow the evolution of the halo from the last merging process until density of 10$^8$ cm$^{-3}$ with the aim to capture the effect of the post-shocked envinronment created by the mergers.
The results obtained are reported in Fig.~\ref{fig:figure2} at different times. As we are interested in examining in detail the effect of the merging process on the chemical and thermal evolution of the simulated halo, we plot the radially averaged mass fractions of the most important species (H$_2$, HD, D$^+$, e$^-$, H$^-$, and H$_2^+$), as well as the temperature and the density profiles. To better understand the chemical behaviour, we also performed a run for an isolated mini-halo with mass roughly equal to the merger case, employing the same chemical network, as well as the same thermal processes. 
The initial conditions for the above isolated halo are taken from our previous work \citep[see halo C in][]{Bovino2013MNRAS,Bovino2014,Grassi2013krome} and the results from this run are also shown in Fig. \ref{fig:figure2}. The comparison is made by taking the data at the same peak density.

\begin{figure*}
\begin{center}
        \includegraphics[width=.3\textwidth]{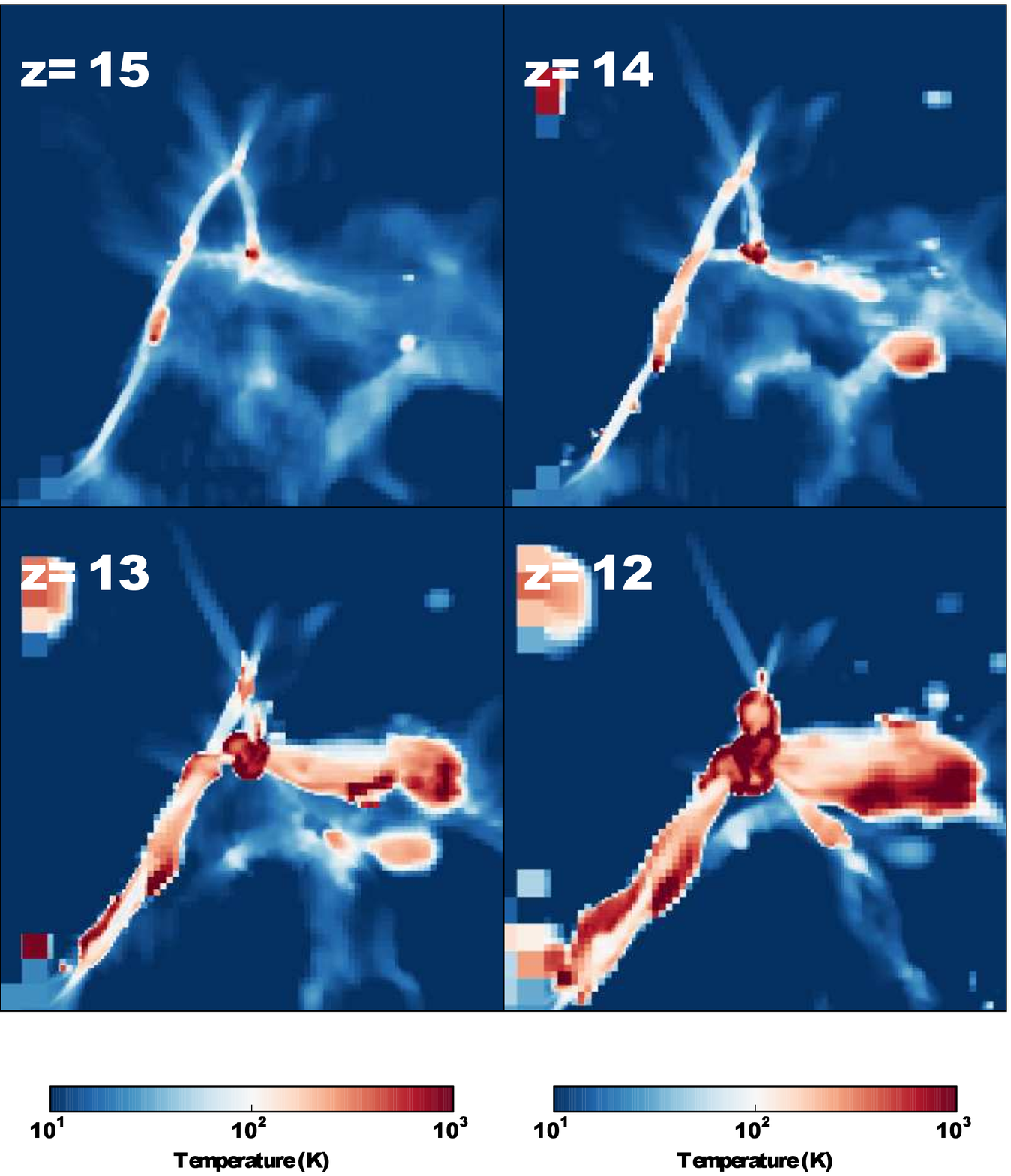}
        \includegraphics[width=.3\textwidth]{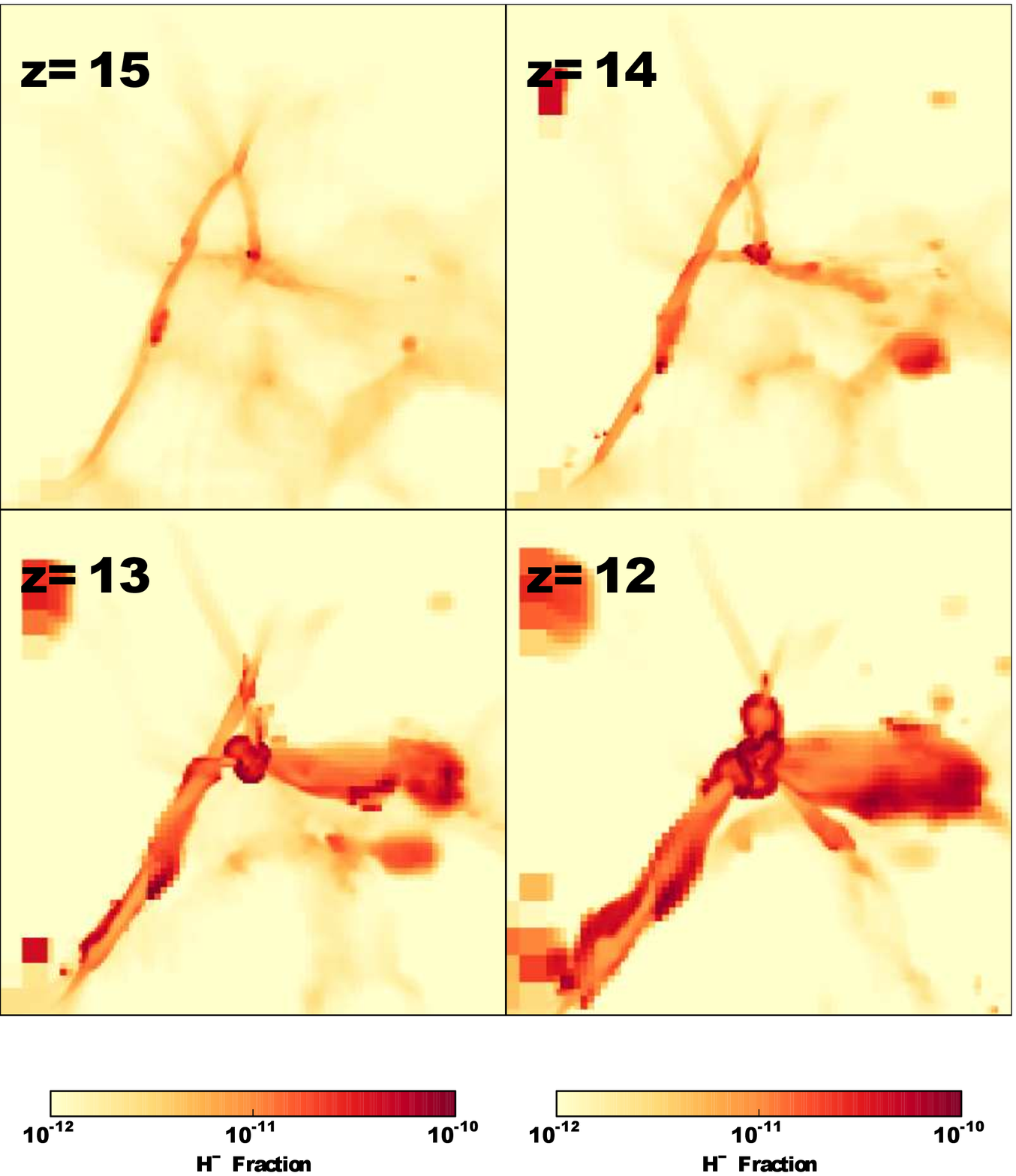}
        \includegraphics[width=.3\textwidth]{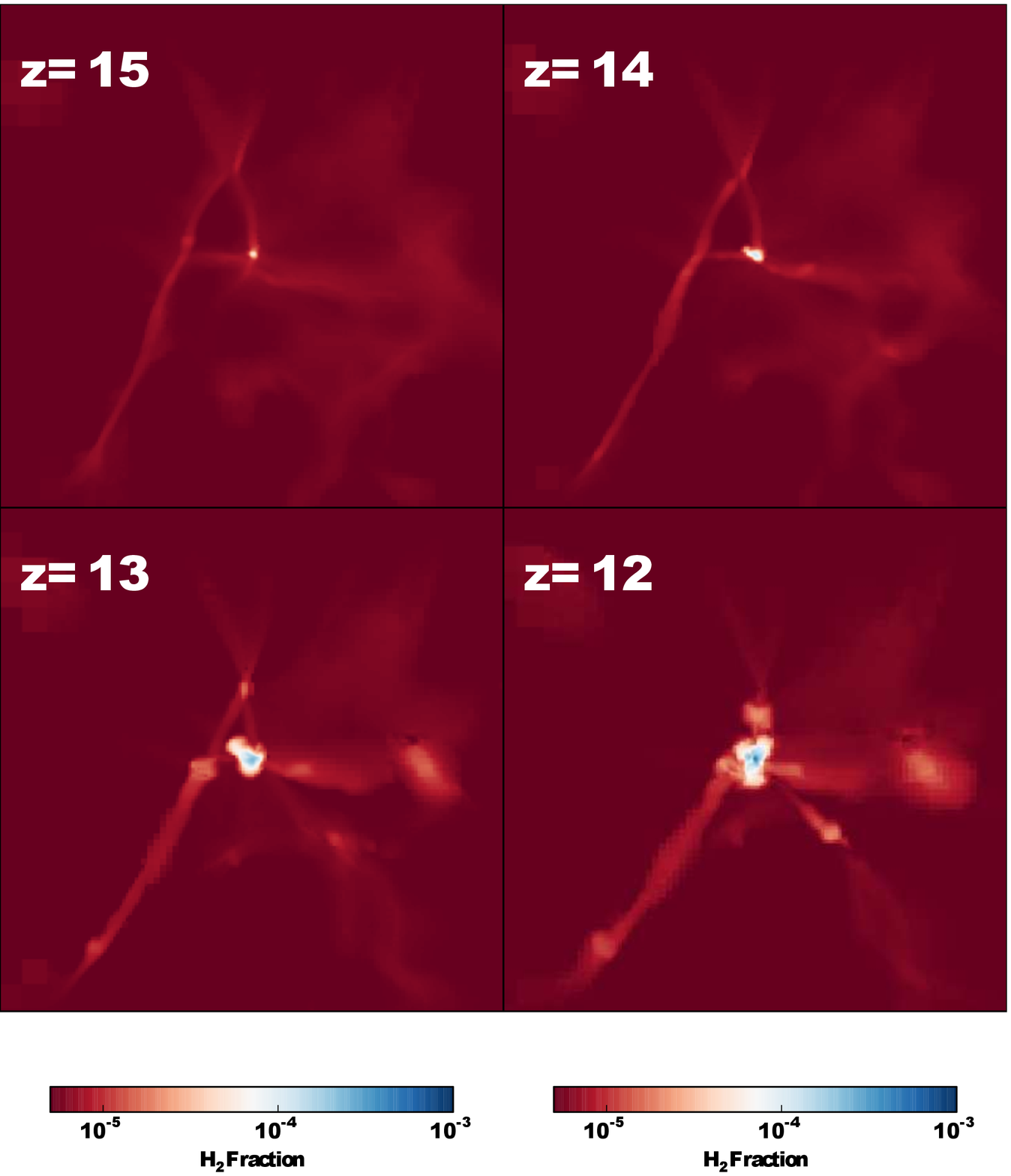}
        \caption{Temperature, H$^-$, and H$_2$ slices on y-direction at different redshifts at a scale of 4 kpc. See text for details.}\label{fig:figure3_1}
\end{center}
\end{figure*}

It is clearly visible from Fig. \ref{fig:figure2} at a scale of $\sim$kpc that a high abundance of H$^-$ is produced by the shocks created during the merging process. This increase in H$^-$ boosts the formation of H$_2$ which is an order of magnitude higher compared to the isolated case. However, the electron fraction is roughly the same for both cases, i.e. of the order of $\sim$10$^{-4}$. To better understand these findings, we show the temperature and local abundances of H$^-$ and H$_2$ in Fig.~\ref{fig:figure3_1} up to a scale of 4 kpc. The presence of merger-induced shocks is clearly visible in the form of enhanced temperatures of above 5000 K. The H$^-$ formation is boosted in the shocked regions, while the H$_2$ fraction is regulated both via H$^{-}$ as well as the underlying density field. The increase of the H$^{-}$ fraction is a consequence of its temperature dependent formation rate (see reaction 7 in table \ref{tab:reactions}). Although no strong differences are observed in the degree of ionization the post-shocked environment created by the mergers triggers the formation of H$^-$ which increases the molecular hydrogen by an order of magnitude (see Fig. \ref{fig:figure2}). Similarly, D$^{+}$ is also enhanced. As the H$_2$ is the main reactant in the formation of HD together with D$^+$ a direct effects is to enhance the overall HD fraction.

As shown in Fig. \ref{fig:figure2}, at smaller radii the difference in the ionization degree is more pronounced. Compared to the isolated halo run the electron fraction is, indeed, almost one order of magnitude higher for radii between 10$^{-2}$ and 1 pc, catalysing the formation of H$_2$ and HD. The core HD fraction is $\sim$10$^{-5}$ in the mergers case compared to $\sim$10$^{-7}$ for the isolated case. We note that HD is gradually increasing over the time and gets boosted in the core because its formation becomes more efficient at lower temperatures and higher densities \citep[see also][]{McGreer2008}. Having HD a hundred times more abundant allows this molecule to cool the gas below 200~K, the temperature where H$_2$ reaches the LTE.
The temperature in the core is about 60 K for the mergers case, while for the isolated halo the temperature remains higher (i.e. $\sim$500 K at the same densities). 
As a results of the lower temperature in the core the electron abundance declines by one order of magnitude in 0.25 Myr (from $z=11.44$ to $z=11.43$) due to recombination processes.

\begin{figure}
\begin{center}
	\includegraphics[width=.5\textwidth]{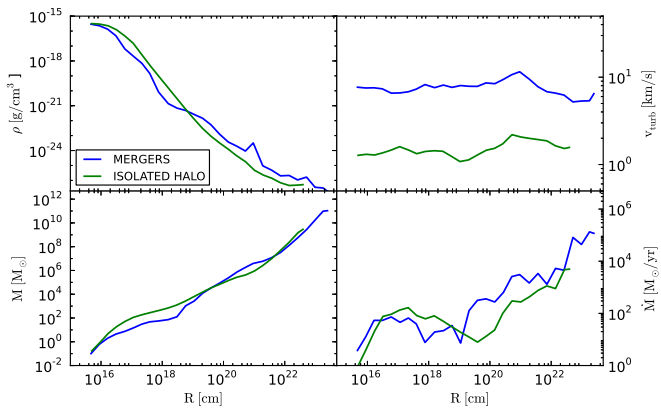}
	\caption{Radially averaged profiles for density (top left), turbulent velocity in km s$^{-1}$ (top right), total mass in solar masses (bottom left), and accretion rate in M$_\odot$ yr$^{-1}$ (bottom right) for the isolated halo (blue dashed) and the mergers run (red solid) for the last dump in the simulation. The comparison data are taken at the same peak density.}\label{fig:figure3}
\end{center}
\end{figure}

\begin{figure*}
\begin{center}
	\includegraphics[width=.8\textwidth]{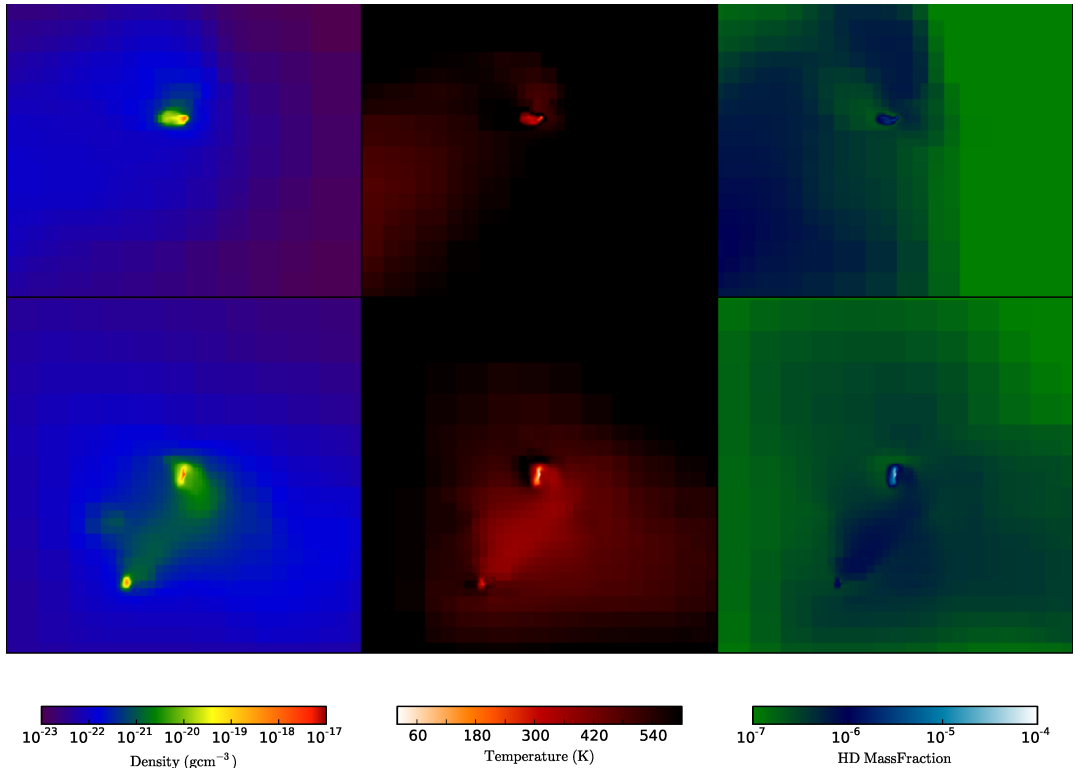}
	\caption{Density, temperature, and HD mass fraction projections, for the run where merging occurs at a scale of 10 pc for two different redshifts, $z=11.44$ (upper panel), and $z=11.43$ (lower panel).}\label{fig:figure4}
\end{center}
\end{figure*}

The dynamical properties of the halo are reported in Fig.~\ref{fig:figure3} for both the merger and the isolated halo. 
Density profiles for both runs are roughly similar and follow an R$^{-2}$ behaviour corresponding to an isothermal collapse, with small deviations for the merger due to the additional fragmentation in the halo.
As expected, the total mass (reported in the bottom left panel) increases as $\sim$R in both cases, and deviations are related to the density profile where shocks are visible for the merger. 
The accretion rate is very large in the outer regions and decreases to 10$^{-1}$~M$_\odot$ yr$^{-1}$ in the core as already reported in previous studies \citep{LatifBH,Latif13}. In the top right panel we report the turbulent velocity in km s$^{-1}$ which is much higher for the merger run. The high turbulent velocity might be due to the presence of the two fragments which induce strong accretion flows.
The halo goes, in fact, through a fragmentation phase as clear from the bottom left panel of Fig.~\ref{fig:figure4} where projections of the density, temperature, and HD mass fraction, at a scale of 10 pc are shown. 

Multiple small clumps are formed in the center of the halo out of which two are visible in Fig. \ref{fig:figure4}. As we are averaging the density along the direction of the projection the third clump is not clearly visible. The clump masses are of about 22, 23 and 0.15 M$_\odot$ and the most massive one is gravitationally bound and has higher rotational support. The central clump shows a very high HD fraction (about 10$^{-4}$) and a temperature of 60 K in the center due to the efficient HD cooling.

The calculations become computationally too expensive at later stages for densities higher than 10$^8$ cm$^{-3}$ and a sink particle approach should be employed to follow the evolution for longer times \citep[e.g.][]{Smith11}. At that point, incorporating feedback from the accretion luminosity would also be relevant. Nevertheless the effect of the HD cooling on the thermal history of the halo is accurately represented if we consider that HD reaches the LTE around 10$^8$ cm$^{-3}$, so that the simulation is well-suited to study the initial fragmentation.

\section{Conclusions and discussion}\label{sect:conclusions}
In this paper we follow the evolution of a mini-halo of $\sim$7$\times$10$^5$ M$_\odot$ formed from the merging of a number of halos with masses of a few 10$^5$ M$_\odot$.
It was in fact suggested \citep{Merging2006} that if two massive enough halos collide and merge the gas can go through a post-shocked ionized stage which boosts the formation of the HD molecule then allowing the gas to
cool below 200 K. As the final mass of the first luminous objects is given by the Jeans mass
\begin{equation}
	M_J = 500\,\mathrm{M_\odot}\left(\frac{T}{200 \mathrm{K}}\right)^{3/2}\left(\frac{10^4 \mathrm{cm^{-3}}}{n}\right)^{1/2}
\end{equation}
it is clear that in an environment dominated by H$_2$ cooling the expected mass is of about 500 M$_\odot$ (assuming $T=200$ K and $n = 10^4$ cm$^{-3}$), while for the HD cooling case the mass is lowered to $\sim$~10 M$_\odot$ (e.g. assuming a temperature of 60~K at $n\sim10^6$ cm$^{-3}$). 

A first attempt to investigate the merger-induced HD cooling was made by \citet{Prieto2012} who performed hydro-cosmological simulations of 3$\times$10$^7$ M$_\odot$ halo resulting from the merging of two minihalos of a few 10$^6$ M$_\odot$. A non-equilibrium treatment of the chemistry was included as well as cooling from H$_2$ and HD. They followed the evolution of the halo until densities of 10$^3$ cm$^{-3}$. The main results from their work was mainly focused on development of turbulence during the merging process. However, they also noticed that few cells of the entire simulations were able to reach temperatures of 100 K with HD fraction around 10$^{-6}$. The densities reached in their work are nevertheless too low to catch possible features due to HD overcooling.

\citet{Prieto2013} continued their investigation studying the statistics of the halo going through a merging process within a cosmological framework but performing DM-only simulations. They suggested that $\sim$30\% of the halos which generates from mergers are able to overcool the gas.

Here we go beyond these studies by coupling the solution of dark matter and hydrodynamics (solved with \verb|ENZO|) with an accurate treatment of the non-equilibrium chemistry (solved with \verb|KROME|) following the collapse until a gas density of 10$^8$ cm$^{-3}$, thus extensively covering the density regime where HD cooling is known to usually dominate \citep{Lipovka2005,GloverAbel08}. We performed high-resolution cosmological simulations for a box of 1 comoving Mpc and follow the evolution of the halo starting from $z=99$, going through the end of the merging process around $z=12$, and following the collapse of our final halo of $\sim$7$\times$10$^5$ M$_\odot$ until $z=11.42$.

Our results clearly show that the shock-waves generated by the merging process enhance H$^-$ and D$^+$ fractions which catalyze the formation of HD. We found a core HD abundance two orders of magnitudes larger than the one obtained with an isolated halo setup. As a consequence, the temperature in the core dropped down to 60-70 K, much lower compared to the isolated halo run where the gas is hotter (around $\sim$500 K).
 
Such over-cooling in the halo has induced fragmentation and multiple clumps are observed, out of which two are rotationally supported and one is gravitationally bound.

It is quite interesting to note that a merging process involving several halos with masses of few 10$^5$~M$_\odot$ can similarly catalyze the required HD fraction via the efficient production of H$^-$. 
 The differences in the masses of the mergers rise from the fact that we performed realistic three-dimensional calculations compared to the simplified one-dimensional estimates by \citet{Merging2006} that provide only a qualitative picture of the process. 
In particular simplified one-dimensional models underestimate the strength of the merger shocks and do not take into account the collapse dynamics and the hydrodynamical effects (e.g. compressional heating). 
Another strong simplification is the assumption of a virial density while realistic halo are partially collapsed, enhancing the formation of molecules. 
Based on the halo finder results from the simulations we found a relative velocity of $\sim$7.0 km s$^{-1}$ between the dark matter halos which is about a factor of 1.5 larger than the estimate by \citet{Merging2006} from their Eq. 2. 

In a previous study, \cite{McGreer2008} found that the fraction of HD-cooled halos is negligible in primordial \emph{unperturbed} environments for halos of masses $\geq$10$^6$~M$_\odot$, but can become relevant for masses $<$10$^6$~M$_\odot$ or in ionized media. The authors explored a series of physical conditions that should favour the formation of HD, i.e. gas efficiently cooled by H$_2$ to a critical temperature which should hold for time long enough to allow the build up of HD molecules. These conditions are more suitable in lower mass halos which collapse more uniformly. On the other hand, \citet{Ripamonti2007} and \citet{Bromm2002} have shown that the critical masses should be of a few 10$^5$~M$_\odot$ and attribute the increase in HD fraction to a longer collapsing time. This was recently confirmed by the comprehensive study by \citet{Hirano2013}. However, in the study by \cite{McGreer2008}, it is not clear if mergers have occurred during the halo formation then a direct comparison with our results is difficult. 
Here, we simulate a situation where the halo formed via the occurance of major mergers and we find that HD abundance is enhanced up to 10$^{-5}$ in the aftermath of the merging process. 
Our findings show that the merger-driven shocks enhance the temperature on large scales, thus favouring the formation of H$^-$
 and increasing the H$_2$ fraction. Also the HD abundance is slightly enhanced at that stage, even though it is unable to
 produce a runaway effect at those densities. When the collapse occurs, the enhanced H$_2$ and HD abundance however allows to trigger a runaway effect as also described by \citet{McGreer2008}. 
Our findings suggest that the fraction of these halos might be larger than previously estimated as merger events are much more common in the early Universe ($\sim$30\% as reported in \citet{Prieto2012}). 

The present work indicates a potential formation route for low-mass primordial stars due to merger-enhanced HD cooling. This pathway should be explored further in future studies, including a larger range of initial conditions.

Finally, the presence of high turbulent energy produced in the aftermath of merging event may also contribute to the amplification of magnetic fields through the so-called small scale dynamo process \citep{Schober2012ApJ,Bovino2013NJP,Schleicher2013AN,LatifBH3,Schleicher2013NJPh,Latif2013MNRAS}. The presence of such fields may further influence fragmentation and disk formation at higher densities \citep[e.g.][]{Machida2013,Latif2013}.

\section*{Acknowledgements}
S.B. and D.R.G.S. thank for funding through the DFG priority program `The Physics of the Interstellar Medium' (project SCHL 1964/1-1). D.R.G.S. and M.L. thank for funding via the SFB 963/1 on "Astrophysical Flow Instabilities and Turbulence" (project A12). The plot of this paper have been obtained by using the $\mathrm{YT}$ tool \citep{Turk2011a}. The simulations have been performed on the Milky Way cluster at the Forschungszentrum J\"ulich. We all are grateful to J. Prieto for having inspired this work and for fruitful discussions.

\bibliographystyle{mn2e}      
\bibliography{mybib_new}

\bsp

\label{lastpage}

\end{document}